\documentclass[usenatbib, useAMS, usegraphicx]{mn2e}

\usepackage{times}
\usepackage[a4paper, totalwidth=480pt, totalheight=680pt]{geometry}

\newcommand{\degree}{\mbox{\ensuremath{^\circ}}}   
\newcommand{\vsini}{\mbox{\ensuremath{v\sin{i}}}}   
\newcommand{\kms}{\mbox{km s$^{-1}$}}   

\title[The spin-orbit alignment of WASP-3b]
{The spin-orbit alignment of the transiting exoplanet WASP-3b from Rossiter-McLaughlin observations\thanks{Based on observations collected with the SOPHIE spectrograph on the 1.93 m telescope at Observatoire de Haute-Provence (CNRS), France, by the SOPHIE Consortium (program 08B.PNP.SIMP).}}

\author[E. K. Simpson et al.]
{E. K.~Simpson$^1$\thanks{Email: esimpson05@qub.ac.uk}, D.~Pollacco$^1$, G.~H\'{e}brard$^2$, N. P.~Gibson$^3$, S. C. C.~Barros$^1$, I.~Boisse$^2$,
\newauthor F.~Bouchy$^{2,4}$, A.~Collier Cameron$^5$, G. R. M.~Miller$^{5}$, C. A.~Watson$^1$  and F. P.~Keenan$^1$\\
$^1$Astrophysics Research Centre, School of Mathematics \& Physics, QueenÕs University, University Road, Belfast BT7 1NN, UK\\
$^2$Institut d'Astrophysique de Paris,  UMR7095 CNRS, Universit\'{e} Pierre \& Marie Curie, 98$^{bis}$ bvd. Arago, 75014 Paris, France\\
$^3$School of Physics, University of Exeter, Stocker Road, Exeter EX4 4QL, UK\\
$^4$Observatoire de Haute-Provence, CNRS/OAMP, 04870 St Michel l'Observatoire, France\\
$^5$School of Physics and Astronomy, University of St Andrews, North Haugh, St Andrews, Fife KY16 9SS, UK}
\date{Submitted to MNRAS 18 Dec 2009, Accepted 20 Feb 2010}

\begin{document}

\maketitle

\begin{abstract}
We present an observation of the Rossiter-McLaughlin effect for the planetary system WASP-3. Radial velocity measurements were made during transit using the \textit{SOPHIE} spectrograph at the 1.93m telescope at Haute-Provence Observatory. The shape of the effect shows that the sky-projected angle between the stellar rotation axis and planetary orbital axis ($\lambda$) is small and consistent with zero within 2$\sigma$; $\lambda=15^{+10}_{-9}$ deg. WASP-3b joins the $\sim$ two-thirds of planets with measured  spin-orbit angles that are well aligned and are thought to have undergone a dynamically-gentle migration process such as planet-disc interactions. We find a systematic effect which leads to an anomalously high determination of the projected stellar rotational velocity ($\vsini=19.6^{+2.2}_{-2.1}$ \kms) compared to the value found from spectroscopic line broadening ($\vsini=13.4\pm1.5$ \kms). This is thought to be caused by a discrepancy in the assumptions made in the extraction and modelling of the data. Using a model developed by \citet{Hirano09} designed to address this issue, we find $\vsini$ to be consistent with the value obtained from spectroscopic broadening measurements ($\vsini=15.7^{+1.4}_{-1.3}$ \kms). 

\end{abstract}


\section{Introduction}\label{Intro}

Bright transiting planets hold the key to determining properties of exoplanet systems which are otherwise inaccessible. One such example is the projected obliquity of the planet's orbit relative to the rotation axis of the star ($\lambda$). This can be measured from spectroscopic observations during transit, which reveal a radial velocity deviation from the standard Keplerian orbital motion known as the Rossiter-McLaughlin (RM) effect \citep{Rossiter24,McLaughlin24}. The effect is caused by the planet passing over and blocking a portion of the rotating stellar surface.  This introduces an asymmetry in the spectral line profile and a small apparent shift in the position of the line. The resulting shape of the RM effect traces the trajectory of the planet across the stellar disc and, once modelled, allows the projected spin-orbit alignment to be found. 

The alignment angle provides an insight into the history of these planetary systems. Hot Jupiters are thought to have formed far out in the system and have migrated inwards. Migration via planet-disk interactions are believed to be relatively non-violent and would not perturb the orbital obliquity \citep{Lin96,Murray98}. It is thought that the timescale for spin-orbit alignment via tidal interactions is of a similar order to the lifetime of the system \citep{Barker09, Levrard09} and therefore that the primordial alignment of the system is preserved. Assuming that a planet was formed in an aligned proto-planetary disc, it too is expected to be well aligned.\footnote{Several authors have noted that this may not necessarily be the case in all systems, especially those which have undergone interactions with other stars, (e.g., \citealt{Golimowski06, Bate09}).} However, more dynamically violent processes such as planet-planet scattering and Kozai oscillations from interactions with a stellar companion are thought to significantly effect the orbital obliquity and could produce highly misaligned and even retrograde orbit \citep[e.g.,][]{RF96,WM03,Nagasawa08}. Thus the distribution of alignment angles allows us to constrain the formation and evolution mechanisms of planetary systems. 

The form of the RM effect depends on $\lambda$ as well as the projected equatorial rotation velocity of the star (\vsini) and several parameters which can be measured from a transit light curve (see Section \ref{sec:params}). For the WASP-3 system, the amplitude of the effect can be estimated as $\sim 0.7 \; \vsini \; (R_{p}/R_{*})^{2} \: \sqrt{1-b^{2}}\sim86$ m s$^{-1}$, where $R_{p}$ and $R_{*}$ are the planetary and stellar radii, $b$ is the impact parameter and 0.7 is a scaling factor which approximates the effects due to limb darkening. WASP-3 is the the fastest rotator for which the RM effect has not been previously reported \citep[$\vsini = 13.4$ \kms, ][]{Pollacco08}. It is also a bright target ($V_{\mathrm{mag}}=10.6$), and therefore an ideal candidate for measuring the RM effect. WASP-3 is a main-sequence star of spectral type F7-8V. It hosts a hot Jupiter of mass 1.8$M_{J}$ and radius 1.3 $R_{J}$ in a close orbit of 1.8 days \citep{Pollacco08,Gibson08}. Due to the close proximity and relative temperature of its host star, WASP-3b is one of the most highly irradiated exoplanets known.     

It has been shown that the formulae used to describe the RM effect (\citealt{Ohta05}) appear to overestimate $\vsini$ compared to the value found from spectral line broadening, and the deviation scales with $\vsini$ \citep[e.g.,][]{Winn05,Triaud09}. This has led \citet{Hirano09} to re-evaluate the equations assuming that the distortion of the radial velocity shift is being measured with a cross-correlation technique (as used in this work). We apply this method to the RM effect of WASP-3 and compare it to the results obtained using the model derived by \citet{Ohta05}. 

In this paper we report on time-series spectroscopic observations of WASP-3 obtained during transit. Section \ref{OandM} describes the observations and analysis procedure performed to measure the spin-orbit angle of the system. The results of the derived system parameters are presented in Section \ref{Results}, and discussed in Section \ref{Conc}. 

\section{Observations and Method} \label{OandM}
 
\subsection{Observations and data reduction}

A transit of WASP-3b was observed with the \textit{SOPHIE} spectrograph at the 1.93m telescope at Haute-Provence Observatory on the night of 2008 September 30. \textit{SOPHIE} is a cross-dispersed, environmentally-stabilised echelle spectrograph (wavelength range 3872.4--6943.5\AA) designed for  high-precision radial velocity measurements \citep{Bouchy09}. The spectrograph was used in high efficiency mode (resolution $R = 40 000$) and the CCD in fast read-out mode. Two 3 arc-second diameter optical fibres were used, the first centred on the target and the second on the sky to simultaneously measure its background in case of contamination from scattered moonlight. The moon illumination was 3\% and at a distance of $>86\degree$ on the night of transit so this did not significantly affect the radial velocity determination.

We acquired 26 spectra of WASP-3 on the night of transit, covering the full transit (137 minutes) and a period of duration 130 minutes out of transit. The exposure times range from 5 to 31 minutes in order to reach a constant signal-to-noise ratio of 35 at 550 nm, reflecting the variable throughputs obtained due to changing atmospheric conditions. The sky was clear when the observations began, and the airmass ranged from 1.1 to 2.7 during the sequence with increasing cloud during the later hours. The exposures remained sufficiently short to provide a good time sampling with 17 measurements taken during transit. In addition, we used 7 out-of-transit observations at various orbital phases obtained during the discovery of the planet in 2007 to fit the orbit \citep{Pollacco08}. 

The spectra were extracted and radial velocities measured using the {\it SOPHIE} pipeline \citep{Perruchot08,Bouchy09}. Radial velocities were computed from a weighted cross correlation of each spectrum with a numerical mask of G2 spectral type, as described by \citet{Baranne96} and \citet{Pepe02}.  A Gaussian was fitted to the cross-correlation functions to obtain the radial velocity shift, FWHM, and contrast with respect to the continuum. The uncertainty in the radial velocity was computed using the empirical relation given in \citet{Bouchy05} and \citet{Cameron07}. Uncertainties were typically $\sim 38$ m s$^{-1}$ during the RM sequence and $\sim 23$ m s$^{-1}$ at other orbital phases.  For the observations taken on the night of transit, the spectrum below 4367 \AA\, was removed from the cross correlation in an effort to reduce the systematic errors caused by the airmass effect. The wavelength stability over the duration of the observations was $\sim3$ m s$^{-1}$. The 7 orbital points reported in \citet{Pollacco08} were re-reduced using an updated version of the pipeline. Table \ref{RVs} summarises the measured radial velocities.

	\begin{table} 
	\centering
	\caption{Radial velocities and 1$\sigma$ error bars of WASP-3 measured with \textit{SOPHIE} during and outside transit. The orbital points are updated from \citet{Pollacco08} using a new version of the reduction pipeline.}
	\begin{tabular}{ccccc}
	\hline \hline
	BJD		&RV		&Error	\\
	-2 400 000	&(\kms)	&(\kms)	\\
	\hline
	Planetary transit:\\
	54740.29509     &  -5.42487        &  0.03856  \\
	54740.29909     &  -5.38158        &  0.03628  \\
	54740.30328     &  -5.36639        &  0.03751  \\
	54740.30771     &  -5.43318        &  0.03734  \\
	54740.31312     &  -5.45738        &  0.03756  \\
	54740.31830     &  -5.40972        &  0.03828  \\
	54740.32351     &  -5.42139        &  0.03816  \\
	54740.32907     &  -5.33800        &  0.03782  \\
	54740.33524     &  -5.35707        &  0.03764  \\
	54740.34152     &  -5.38229        &  0.03741  \\
	54740.34699     &  -5.38596        &  0.03626  \\
	54740.35300     &  -5.36107        &  0.03533  \\
	54740.35891     &  -5.36891        &  0.03508  \\
	54740.36506     &  -5.41906        &  0.03437  \\
	54740.37170     &  -5.49966        &  0.03580  \\
	54740.37771     &  -5.55305        &  0.03686  \\
	54740.38426     &  -5.55488        &  0.03571  \\
	54740.39228     &  -5.58417        &  0.03825  \\
	54740.39993     &  -5.64022        &  0.03606  \\
	54740.40758     &  -5.58523        &  0.03519  \\
	54740.41466     &  -5.69115        &  0.03415  \\
	54740.42259     &  -5.64015        &  0.03543  \\
	54740.43066     &  -5.53109        &  0.03945  \\
	54740.44171     &  -5.59288        &  0.03844  \\
	54740.45894     &  -5.62403        &  0.04132  \\
	54740.48062     &  -5.52085        &  0.05355  \\
	\hline
	Other orbital phases:\\
	54286.52255    & -5.72553      &  0.02137 \\
	54287.45639    & -5.18430      &  0.03541\\
	54289.36622    & -5.18792      &  0.03047\\
	54340.32511    & -5.64313      &  0.01280\\
 	54341.39898    & -5.40394      &  0.01914\\
	54342.31981    & -5.54401      &  0.02801\\
	54343.48259    & -5.60354      &  0.01283\\
	\hline
	\end{tabular}
	\label{RVs}
	\end{table}
	
\subsection{Determination of system parameters}\label{sec:params}
	
The RM effect and orbit were fitted simultaneously using all the available spectroscopic data. A Keplerian model was used for the orbit, and the analytical approach described in \citet{Ohta05} (hereafter OTS) for the RM effect. The OTS equations were modified to make them dependent on $R_{p}/R_{*}$ and $a/R_{*}$ rather than $R_{p}$, $R_{*}$ and $a$, to reflect the parameters derived from photometry, and reduce the number of free parameters. The model comprises 13 parameters: orbital period, $P$; mid-transit time, $T_\mathrm{0}$; planetary to stellar radius ratio, $R_{p}/R_{*}$; scaled semi-major axis $a/R_*$;  orbital inclination, $i$; orbital eccentricity, $e$; longitude of periastron, $\omega$; radial velocity semi-amplitude, $K$; systemic velocity of orbital dataset, $\gamma_{1}$;  systemic velocity of transit dataset, $\gamma_{2}$; sky projected angle between the stellar rotation axis and orbital angular momentum vector, $\lambda$; projected stellar rotational velocity, $v \sin {i} $; and stellar linear limb-darkening coefficient, $u$. 

The six observations taken immediately before the transit ingress indicate that the systemic velocity of the transit data-set is $\sim25$ m s$^{-1}$ lower than that measured from the orbital velocities. We therefore allow the systemic velocity of the orbit and transit data to differ $(\gamma_{1}$ and  $\gamma_{2})$. All of the out of transit points (from 2007 and 2008) were first fitted and no significant deviation from a circular orbit was found ($e=0.07\pm0.08$). The eccentricity was then fixed to zero for the rest of the analysis. 

We used a linear limb darkening law, as the quadratic law alters the model by less than $\sim3$ m s$^{-1}$ and so does not seem justified by the precision of the RM data, especially given that for $ b\sim 0.5$ as in this case, the alignment angle is primarily determined by the time when the radial velocity anomaly vanishes, rather than the exact form of the model. The linear limb-darkening coefficient $u = 0.69$ was chosen from the tables of \citet{Claret04} (ATLAS models) for the $g'$ filter . We used the value corresponding to $T_{\mathrm{eff}} = 6500$\,K, $\log g=4.5$, $[\mathrm{M/H}]=0.0$ and $\zeta =2$ \kms (a typical value for this spectral type, see \citealt{Gray01})  which most closely represented the properties of WASP-3 ($T_{\mathrm{eff}} = 6400$\,K, $\log g=4.25$, $[\mathrm{M/H}]=0.0$, \citealt{Pollacco08}). As a test, we set $u$ as free parameter and found that there was no significant effect on $\lambda$ or $\vsini$. 

We placed no constraint on $\vsini$ as there is a known systematic effect in its determination from RM measurements (see Section \ref{vsini}). \citet{Triaud09} suggest that there may be a similar systematic effect influencing $R_{p}/R_{*}$ if solely spectroscopic observations are used.\footnote{This is because the amplitude of the RM effect is primarily determined by $R_{p}/R_{*} $ and $\vsini$, so the observed increased amplitude could lead to a false inflation of these values.} We therefore fixed the photometric parameters ($P$, $T_0$, $R_p/R_*$, $a/R_{*}$ and $i$) to those found through light curve modelling. An MCMC fit to a full and partial transit of WASP-3 was presented in \citet{Gibson08}. In this analysis, the stellar radius was calculated based on the scaling relation $R_{*} \propto M_{*}^{1/3}$ with a prior on $M_{*}$, leading to an underestimation in the error on the orbital inclination. The light curves were re-fitted with the stellar radius as a free parameter using the procedure described in \citet{Gibson09}. Five separate chains each with two-hundred thousand points were computed. To test that the chains had all converged to the same parameter space, the Gelman \& Rubin statistic \citep{GR92} was calculated for each of the free parameters. It was found to be less than $1\%$ from unity for all parameters, a good sign of mixing and convergence. The results are shown in Table \ref{Adopted}. 

	\begin{table}
	\centering
	\caption{Adopted system parameters and uncertainties required for the RM effect, and other photometric parameters updated from \citet{Gibson08}.}	
	\label{Adopted}
	\begin{tabular}{lc r@{}l}
	\hline\hline
	Parameter (units)			&Symbol & \multicolumn{2}{c}{Value}\\
	\hline
	Period (days) 				&$P$				&$1.846835~$			&$\pm~0.000002$			\\[4pt]
	Transit epoch	(HJD) 		&$T_\mathrm{0}$		&$2454605.55915$		&$\pm~0.00023$			 \\[4pt]
	Orbital inclination (deg)		& $i$ 				&$84.93\,$			&${}^{+1.32}_{-0.80}$			\\[4pt]
	Planet/Star radius ratio		& $R_{p}/R_{*}$		&$0.1013\,$			&${}^{+0.0014}_{-0.0013}$		\\[4pt]
	Scaled semi-major axis		&$a/R_{*}$ 			&$5.173\,$			&${}^{+0.246}_{-0.162}$		\\[4pt]
	Eccentricity $^{*}$			&$e$				&$0\,$ 				&\,(adopted)				\\[4pt]
	Limb darkening $^{\dagger}$	&$u\;$				&$0.69\,$				&						 \\
	\hline
	Impact parameter 			&$b$				&$0.466\,$			&${}^{+0.047}_{-0.111}$		\\[4pt]
	Transit duration (hours)		&$T_{d}$				&$2.737\,$				&${}^{+0.024}_{-0.022}$		\\[4pt]
	Stellar radius ($R_{\odot}$)	&$R_{*}$				&$1.31\,$				&${}^{+0.05}_{-0.07}$		\\[4pt]
	Planet radius ($R_{J}$)		&$R_{p}$				&$1.29\,$				&${}^{+0.05}_{-0.07}$		\\
	\hline
	\multicolumn{3}{l}{$^{*}$ See Section \ref{OandM} }\\
	\multicolumn{3}{l}{$^{\dagger}$ from \citet{Claret04}}
	\end{tabular}
	\end{table}

Thus the model contains 5 free parameters ($\lambda$, $v \sin{i}$, $K$, $ \gamma_{1}$ and $\gamma_2$). The $\chi^2$ statistic was calculated: 

\begin{equation}
	\chi^{2} = \sum_{i}^{} \left [ \frac{v_{i, \mathrm{obs}} - v_{i, \mathrm{calc}}}{\sigma_{i}} \right ]^{2}
\end{equation}

\noindent where $v_{i, \mathrm{obs}}$ and $v_{i, \mathrm{calc}}$ are the $i$th observed and calculated radial velocities and $\sigma_{i}$ the corresponding observational error. The optimal parameters were obtained by minimising the $\chi^2$ statistic using the IDL-based MPFIT function \citep{mpfit}; a least-squares minimisation technique using the Levenberg-Marquardt algorithm. The starting values for the parameters were those derived in \citet{Pollacco08} as well as $\lambda = 0$ and $\gamma_2 = \gamma_{1}$.

The spectroscopic errors were rescaled so that the best fitting model has a reduced $\chi^{2}$ ($\chi^{2}_{red}$) of 1, which required the initial spectroscopic errors to be multiplied by the factor $\sqrt{\chi^{2}_{red}}=0.905$. This suggests that the errors were slightly overestimated however the rescaling did not significantly affect the best-fit parameters or errors.  

The $1\sigma$ best-fit parameter uncertainties were calculated using a Monte-Carlo method.  We created $10^{5}$ synthetic data sets by adding a 1$\sigma$ Gaussian random variable to the data points. The photometric parameters were also varied for each realisation by taking a parameter set from the distributions from the light curve fitting. This allowed correlations between parameters such as $a/R_{*}$ and $i$ to be taken in to account. The starting values for the free parameters were also perturbed by a 3$\sigma$ Gaussian random variable to further explore the parameter space. The free parameters were re-optimised for each simulated data-set to obtain the distribution of the best-fit parameter values. The distributions were not assumed to be Gaussian and the 1$\sigma$ limits were found where the distribution enclosed $\pm34.1\%$ of the values away from the median. In order to verify the errors, we also ran a global fit to both the photometric and radial velocities using an MCMC algorithm, as described in \citep{Anderson09}. 

	\begin{figure} 
	\centering
	\includegraphics[angle=0, width=\linewidth]{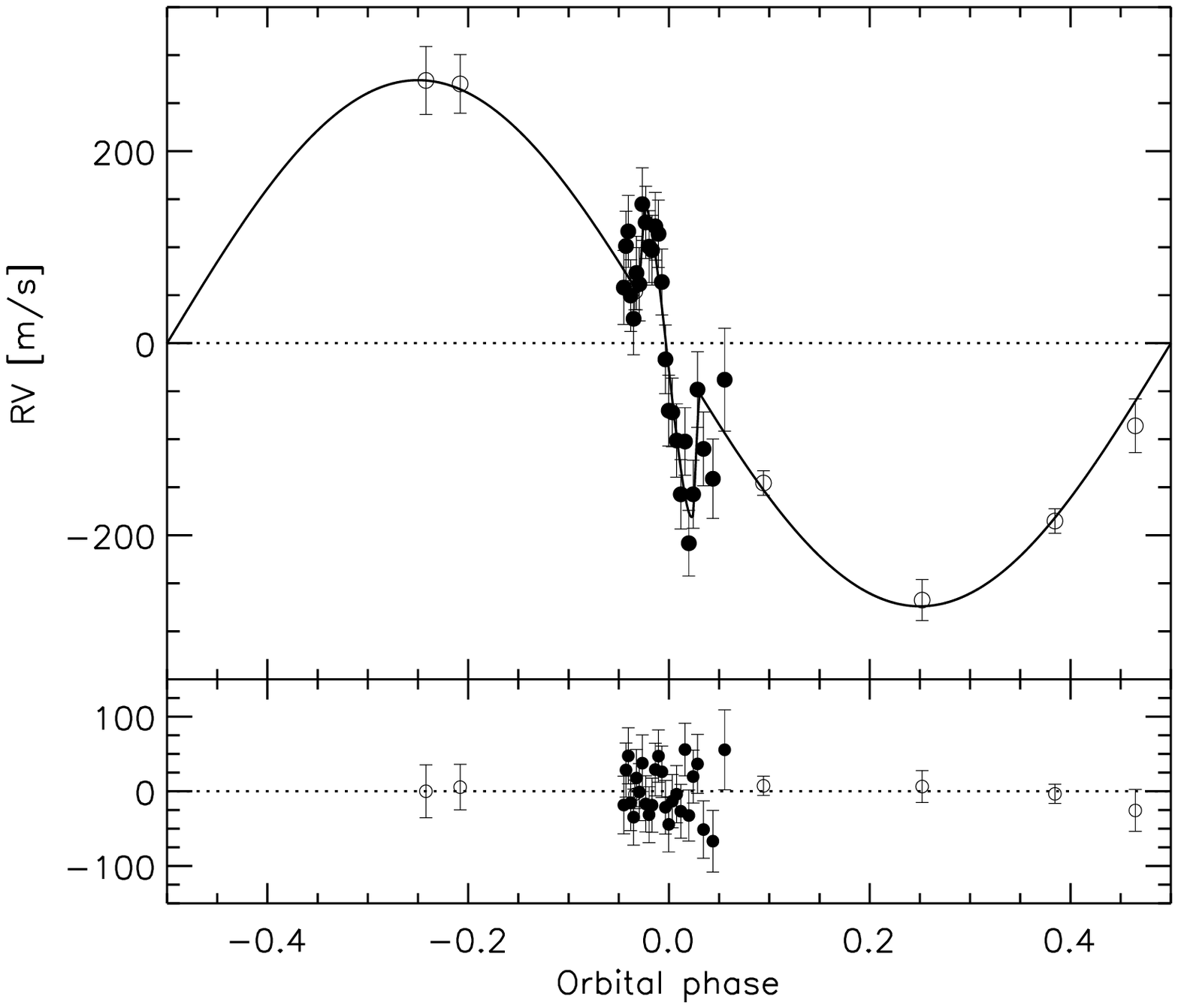}	
	\caption{Upper panel: Radial velocities of WASP-3 minus the systemic velocity, overplotted with the best-fit model. Open points are from \citet{Pollacco08}. Filled points show the RM effect of WASP-3 on 2008 September 30. Lower panel: residuals from the best-fit curve.}
	\label{fig:orbit}
	\end{figure}
	
	\begin{figure}    
	\centering
	\includegraphics[angle=0, width=\linewidth]{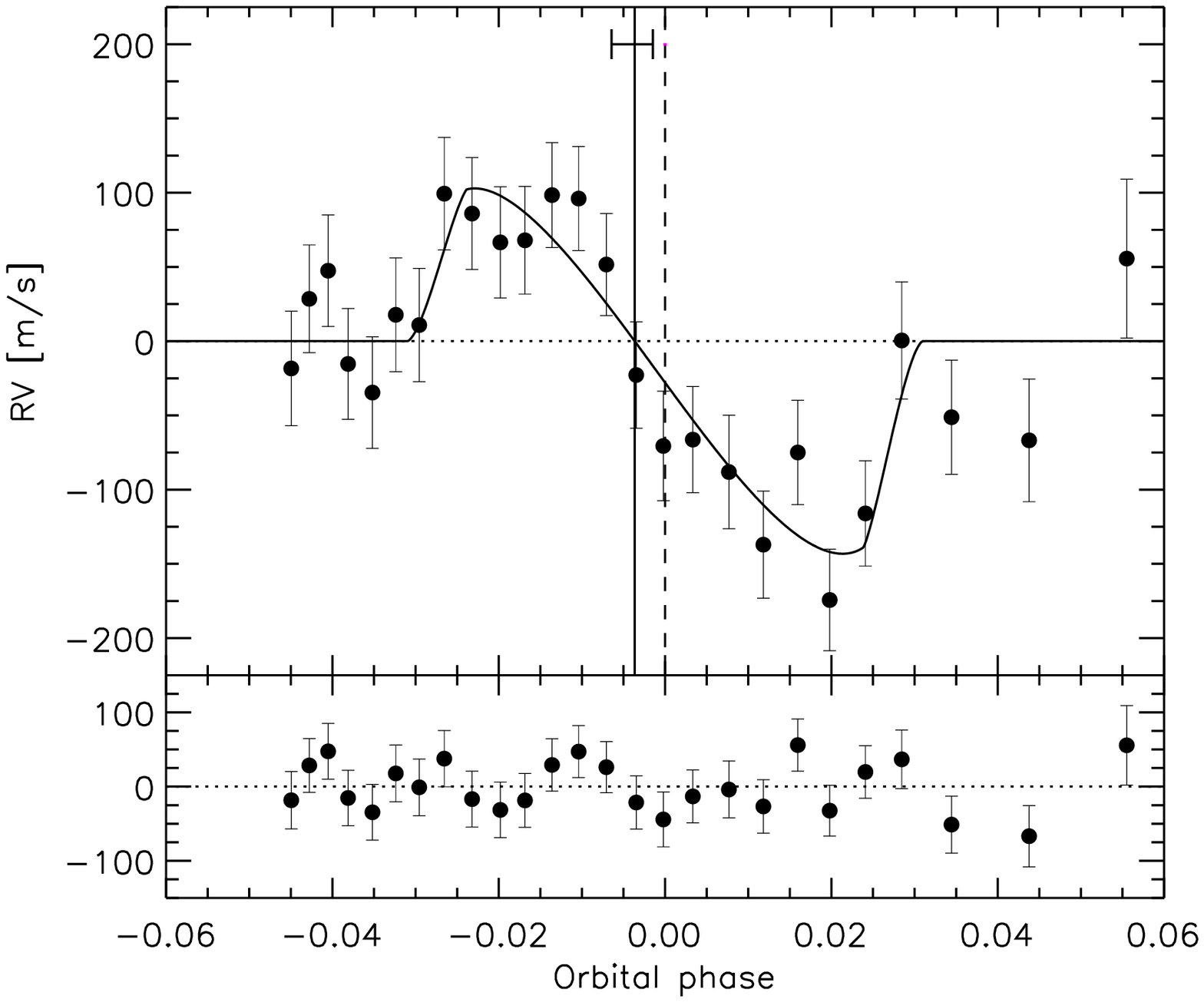}
	\caption{Upper panel: RM effect of WASP-3 minus the Keplerian motion, overplotted with the best-fit model. The solid and dotted lines show the offset between the radial velocity and photometric zero-points, suggesting that the system is not completely aligned. The error bar on the solid line shows the 1$\sigma$ uncertainty in the position of the radial velocity zero-point. The uncertainty on the photometric zero-point is smaller than the width of the line. Lower panel: residuals from the best-fit model.}
	\label{fig:transit}
	\end{figure}
	
	\begin{table*}
	\centering
	\caption{Best-fit values and uncertainties using \citet{Ohta05} (OTS) (both Least-squares and MCMC fitting) and \citet{Hirano09} (H09) models.}
	\begin{tabular}{lc r@{}l r@{}l r@{}l}
	\hline\hline
	Parameter (units)		&Symbol 	& \multicolumn{2}{c}{OTS}&  \multicolumn{2}{c}{OTS (MCMC)} &  \multicolumn{2}{c}{H09} \\
	\hline
	Projected alignment angle (deg)		&$\lambda$		&$15\,$		&${}^{+10}_{-9}	$	&$14~$		&$\pm~9$		&$13\,$		&${}^{+9}_{-7}$	\\[4pt]
	Projected stellar rotation velocity (\kms)	& $v\sin{i}$ 		&$19.6\,$		&${}^{+2.2}_{-2.1}$	&$19.6~$		&$\pm~2.2$	&$15.7\,$		&${}^{+1.4}_{-1.3}$\\[4pt]
	RV semi-amplitude 	(\kms)			& $K$ 			&$0.274~$	&$\pm~0.011$		&$0.273~$	&$\pm~0.013$	&$0.276~$	&$\pm~0.011$\\[4pt]
	Systemic velocity of orbit dataset (\kms)	& $\gamma_1$ 	&$-5.458~$	&$\pm~0.008$		&$-5.458~$	&$\pm~0.005$	&$-5.458~$	&$\pm~0.007$	\\[4pt]
	Systemic velocity of transit dataset (\kms)	& $\gamma_2$		&$-5.483~$	&$\pm~0.011$		&$-5.486~$	&$\pm~0.011$ 	&$-5.487~$	&$\pm~0.009$	\\[4pt]
	Chi-squared (spectroscopic data)		& $\chi ^{2}$		& \multicolumn{2}{c}{27.9}		& \multicolumn{2}{c}{27.4}	&  \multicolumn{2}{c}{27.2}	\\	
	\hline
	\end{tabular}
	\label{Output}
	\end{table*}

\section{Results} \label{Results}

\subsection{System parameters}

The derived parameters and errors for the WASP-3 system are shown in Table \ref{Output} for both the least-squares and MCMC methods. The results from the global MCMC analysis are consistent with those found from the least squares fitting. The re-reduced radial velocities yield orbital parameters which are consistent with \citet{Pollacco08}.

Figure \ref{fig:orbit} shows all the WASP-3 radial velocities minus the systemic velocity; the best-fit model; and the residuals to the fit. Figure  \ref{fig:transit} shows a zoom-in on the observations taken on the night of transit minus the Keplerian orbit; the best fit model; and residuals. It can be seen that there is a $\sim10$ minute offset between the radial velocity and photometric zero points (solid and dotted lines). The elapsed time between the transit photometry and the RM observations is approximately 73 orbits and the 0.173s uncertainty in $P$ and 20s uncertainty in $T_{0}$ causes an uncertainty of only 33 s which cannot explain the offset. By fitting the data, we obtain $\lambda=15^{+10}_{-9}$ deg, indicating that the timing offset is caused by a small misalignment between the planetary orbit and stellar spin axis. The system is generally well aligned and the misalignment is not highly significant, with $\lambda$ consistent with zero to within $2\sigma$.

The difference between the offset velocities of the 7 orbital data points and those taken on the night of transit is significant to 2 $\sigma$; $\sim28 \pm 14$ ms$^{-1}$, so we investigated modelling all the data together without including a relative shift. We found that the newly derived parameters were consistent with the previous values to within the error bars. The value of $\lambda$ increased to 23$^{+8}_{-7}$ deg because $\lambda$ is highly correlated with $\gamma$. The new fit yields a $\gamma$ velocity which is higher than that found for the transit dataset alone, so the RM effect appears to cross the zero velocity point earlier, and produces a larger misalignment angle. There are several factors which could cause an offset between the datasets. These include a long-term trend caused by an additional body, stellar activity, instrumental drift or a combination of these.\footnote{The shift is larger than the typical value for astrophysical jitter for this spectral type so this is unlikely to be the sole cause.} In addition, the data taken on the night of transit underwent a slightly different reduction procedure which may well cause a shift in the radial velocities between nights, (see Section \ref{OandM}) so we prefer to fit the two datasets separately.

\subsection{Stellar rotational velocity and systematics} \label{vsini}

\begin{figure} 
	\centering
	\includegraphics[angle=0, width=\linewidth]{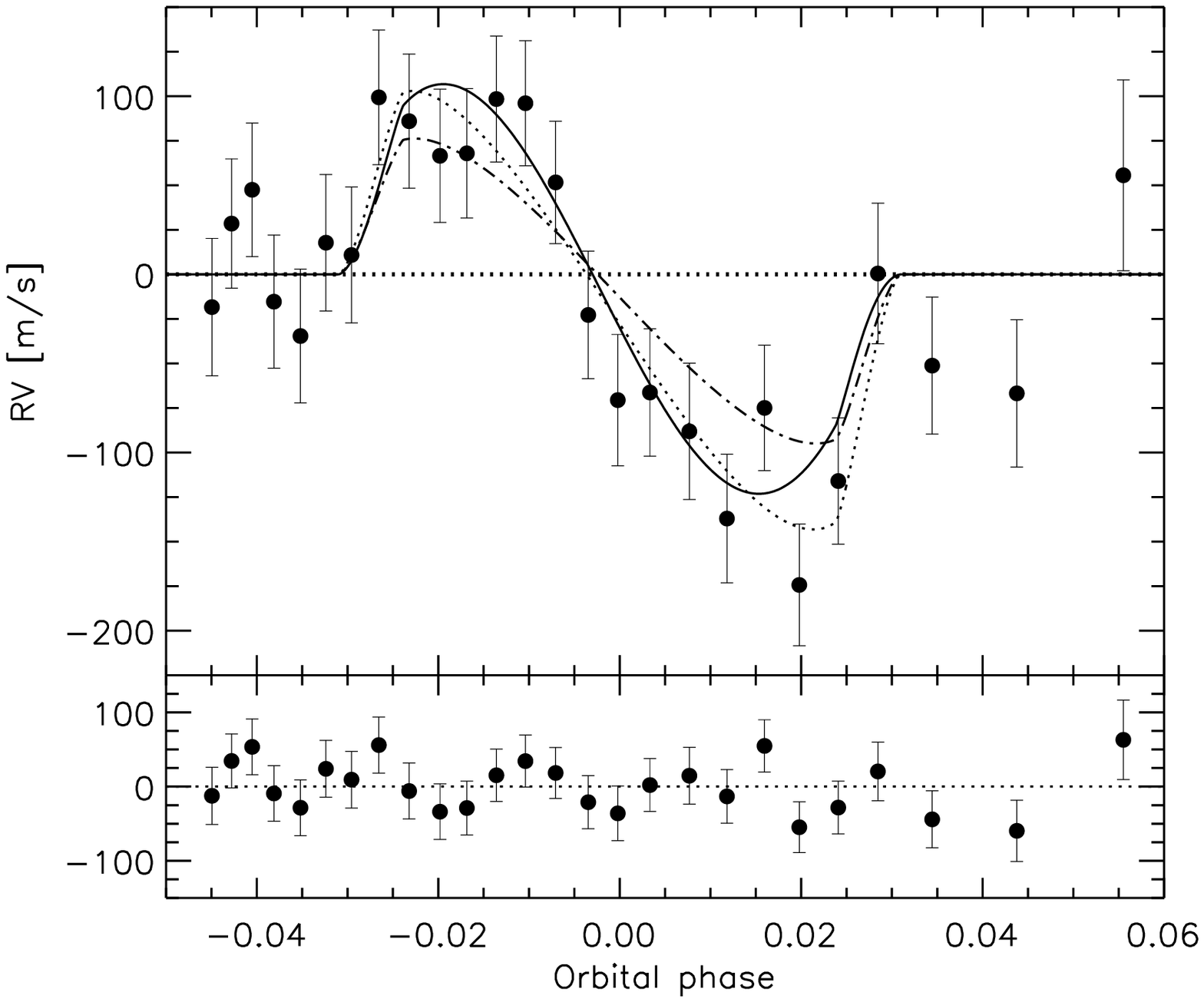}
	\caption{Upper panel: RM effect of WASP-3 with various models. Solid (red) is the fit using the \citet{Hirano09} (H09) model with $\vsini$ as a free parameter ($\vsini=15.3$ \kms). Dotted (blue) is the \citet{Ohta05} (OTS) model with $\vsini$ as a free parameter ($\vsini=19.3$ \kms). Dash dot (blue) is the fit using the OTS model but with $\vsini$ held fixed at $\vsini=13.4$ \kms as determined through spectroscopic broadening \citep{Pollacco08}. The amplitude of this model is obviously too low. All the models cross the x-axis at the same phase, giving consistent values of $\lambda$. Lower panel: residuals from the H09 best-fit model.}
	\label{RMfast}
	\end{figure}

The observed amplitude of the RM effect ($\sim100$ m s$^{-1}$) is significantly larger than expected ($\sim86$ m s$^{-1}$, see Section \ref{Intro}). Figure \ref{RMfast} shows the predicted RM effect with $\vsini$ equal to that found from modelling the spectral line broadening ($\vsini = 13.4\pm1.5$ \kms; \citealt{Pollacco08}).  By fitting the RM effect, we find the projected rotational velocity to be overestimated by $\sim50\%$; $\vsini=19.6^{+2.2}_{-2.1}$ $\kms$. Several authors have noted a similar discrepancy, which is particularly evident in faster rotating stars. \citet{Winn05} suggested that systematics are present in the determination of the radial velocities because the asymmetric stellar line profile is being fitted with a symmetric Gaussian. They attempted to correct this by adding a quadratic term to the RM model which significantly improved the estimated $\vsini$.

\citet{Triaud09} showed that in HD 189733, the structure of the residuals observed when subtracting the RM model from the data was consistent with those found when subtracting the RM model from simulated data undergoing the same data reduction procedure. This indicates that the cause of the discrepancy is indeed in the extraction of the radial velocities. The residuals of the WASP-3 RM effect show a very similar structure during the first three quarters of the transit to those in Figure 4 of \citet{Triaud09} (the last section is dominated by systematics caused by airmass and weather effects). The amplitude of the residuals is of the order of the error bars so it is not clear whether the structure is detected significantly. 

The apparent radial velocity shift of the stellar lines predicted by the OTS model is produced by the distortion of the line profiles by the planet blocking of a velocity component of the star's spectrum. Assuming that the resulting asymmetry of the stellar lines is not well resolved, a symmetric function fitted to the line profile measures a shift in the mean position. For faster rotating stars, the asymmetry in the line profile is better resolved because of the broader width, and causes a larger apparent shift than expected, as seen here.

\citet{Hirano09} (hereafter H09) have addressed this problem by modifying the OTS equations under the assumption that the RV shift is measured by cross-correlation with a Gaussian function. They find that the RM model is also dependent on the intrinsic width of the individual line profile. 

Hence, instead of modelling the RM velocity anomaly using the OTS equation;
\begin{equation} 
 \delta v = f v_{p}/(1-f)
\end{equation}

\noindent where $f$ is flux blocked by planet $\sim(R_{p}/R_{*})^{2}$ and $v_{p}$ is sub-planet velocity (ie the velocity component of the rotating stellar surface blocked by the planet $\sim x_{p} ~ \vsini ~ /R_{*} $, where $ x_{p}$ is the x co-ordinate of the position of the planet on the stellar surface, see Fig 5 of OTS), we now use the H09 model (Equations 41 and 42): 

\begin{equation} 
	\delta v = f  v_{p}  (p - qv_{p}^{2}) = f v_{p} (1.72 - 0.00546 ~ v_{p}^{2})
\end{equation} 

\noindent where
 	
\begin{equation} 
	p = \left [1 + \frac{\sigma^{2}}{2\beta^{2}+\sigma^{2}}\right]^{3/2}, \hspace{20pt} q = \frac{p}{2\beta^{2}+\sigma^{2}}
\end{equation} 

\noindent and $\beta$ is the intrinsic line width, which has been calculated by Miller et al. (in prep) for WASP-3 as $\beta =  11.1 \pm 0.1 \kms$. The quantity $\sigma=\vsini/\alpha$, where $\alpha$ is a scaling factor depending on limb darkening parameters (see Equation F6 of H09). We calculate $\alpha=1.31$ for the fixed limb darkening coefficients $u_{1}=0.69$ and $u_{2}=0$. 

We fit this equation to the data and find that $\lambda$ is consistent with that found using the OTS model, $\lambda = 13^{+9}_{-7}$ deg  ($\lambda = 15^{+10}_{-9}$ deg; OTS). This is expected because $\lambda$ mainly depends on the shape of the waveform and the time crossing zero point rather than the amplitude. The slight difference is probably caused by fit finding a marginally lower value of $\gamma_{2}$, which is correlated with $\lambda$. We find $\vsini$ to be much closer to the line-broadening value $\vsini = 15.3 \pm 1.4$ \kms. It is still slightly different but is acceptable considering the errors. There was no significant difference in $\chi^{2}$ between the OTS and H09 fits.  Figure \ref{RMfast} shows the data over-plotted with the best-fit OTS and H09 models.

We also investigated whether the discrepancy between the two independent measurements of $\vsini$ could be caused by differential rotation. \citet{GW07} show that if the surface speed of a star is dependant on latitude as well as longitude, then the form of the RM effect is slightly altered and depends on the fractional difference in rotation speed between the pole and equator ($\alpha$). To reproduce the $\sim$6 \kms difference in $\vsini$ seen in WASP-3, $\alpha$ must be greater than 0.45. \citet{Reiners06} showed that no stars have been observed with $\alpha > 0.45$ so this scenario seems unlikely. In addition, the form for the differential rotation law would have to be anti-solar, which is seldom observed. The bias caused by the fitting of gaussians to asymmetric CCFs offers a much more viable explanation.

\section{Conclusions} \label{Conc}
A full transit of WASP-3b was observed with the \textit{SOPHIE} spectrograph at the 1.93m telescope at Haute-Provence Observatory. We modelled the RM effect using the equations of \citet{Ohta05} and found the sky projected spin-orbit alignment angle of the system to be $\lambda=15^{+10}_{-9}$ deg. WASP-3b is not strongly misaligned and is consistent with $0\degree$ to within $2\sigma$. This suggests that WASP-3b underwent a relatively non-violent migration process which did not perturb it from the primordial alignment of the proto-planetary disc. The result indicates a marginal detection of a small misalignment, and further, higher precision observations of this target would establish the reality of this. 

We find a significant discrepancy between the $\vsini$ value from the RM model ($\vsini=19.6^{+2.2}_{-2.1}$ \kms) and the measurement from spectroscopic line broadening analysis ($\vsini = 13.4\pm1.5$ \kms; \citealt{Pollacco08}). This phenomenon has been noted in several systems, especially for fast rotators. It is thought to be the result of systematics introduced when measuring a line profile asymmetry as a shift in the mean line position. 

\citet{Hirano09} provide a modified formulation to take into account the method used to determine radial velocity shifts. We applied this model and found the alignment angle to be consistent with the OTS model, and derived an improved value for the rotational velocity $\vsini=15.7^{+1.4}_{-1.3}$ \kms. This is consistent with the value found from spectroscopic line broadening analysis. The RM effect has the potential to  provide a much more precise determination of $\vsini$ than spectral broadening for slow rotators, as it is a geometrical measurement that is not subject to uncertainties such as macroturbulence. However, RM models must be shown to ensure that the $\vsini$ discrepancy is fully resolved.

Since the beginning of 2009, the number of systems reported with tilts significantly greater than $30\degree$ has risen dramatically from one \citep[XO3,][]{Hebrard08} to six (CoRoT-1, \citealt{Pont09a}; HAT-P-7, \citealt{Winn09a,Narita09}; HD 80606, \citealt{Moutou09, Pont09b, Gillon09, Winn09b}; WASP-14, \citealt{Johnson09}; WASP-17, \citealt{Anderson09}). The RM effect has now been measured for a total of 18 systems and WASP-3 joins the $\sim2/3$ of these which are well aligned. These statistics are re-enforcing the inference by \citet{FW09}  that there is a bimodal distribution of spin-orbit angles, with a fraction of systems being well aligned while the rest have random mutual orientations, suggesting that there are multiple migration mechanisms at work. It is therefore vital to increase the number of measured systems, both aligned and misaligned, to place these theories on a secure statistical basis.

\section*{Acknowledgments}

We thank the technical team at Haute-Provence Observatory for their support with the \textit{SOPHIE} instrument and the 1.93-m 
OHP telescope. FPK is grateful to AWE Aldermaston for the award of a William Penny Fellowship. EKS would like to thank T. Hirano for his help with the modified RM equations and R.M. Crockett for his support and nursing of injuries during the observing campaign.

\bibliographystyle{mn2e.bst}
\bibliography{bib.bib}

\end{document}